\documentclass[acmsmall,screen]{acmart}

\usepackage{xspace}
\usepackage{multirow}
\usepackage{array}

\newcolumntype{?}{!{\vrule width 1pt}}

\newcommand{\tool}{ChatGPT\xspace}
\newcommand{\taskCount}{48\xspace}
\newcommand{\taskGPT}{22\xspace}
\newcommand{\taskAuto}{13\xspace}
\newcommand{\taskManual}{35\xspace}
\newcommand{\qCount}{32\xspace}
\newcommand{\cBetter}{97\xspace}
\newcommand{\codefont}[1]{\footnotesize{\texttt{#1}}\normalsize}
\newcommand{\repoCount}{357\xspace}

\newcolumntype{L}[1]{>{\raggedright\let\newline\\\arraybackslash\hspace{0pt}}m{#1}}
\newcolumntype{C}[1]{>{\centering\let\newline\\\arraybackslash\hspace{0pt}}m{#1}}
\newcolumntype{R}[1]{>{\raggedleft\let\newline\\\arraybackslash\hspace{0pt}}m{#1}}
\AtBeginDocument{%
  }

\setcopyright{acmcopyright}
\copyrightyear{2018}
\acmYear{2018}
\acmDOI{XXXXXXX.XXXXXXX}

\acmPrice{15.00}
\acmISBN{978-1-4503-XXXX-X/18/06}

\begin{document}

\title{An empirical study of ChatGPT-3.5 on question answering and code maintenance}

\author{Md Mahir Asef Kabir}
\affiliation{%
  \institution{Virginia Tech}
  \country{USA}}
\email{mdmahirasefk@vt.edu}

\author{Sk Adnan Hassan}
\affiliation{%
  \institution{Walmart, Inc.}
  \country{USA}}
\email{adnan.hassan@walmart.com}

\author{Xiaoyin Wang}
\affiliation{%
  \institution{University of Texas at San Antonio}
  \country{USA}}
\email{xiaoyin.wang@utsa.edu}

\author{Ying Wang}
\affiliation{%
  \institution{Northeastern University}
  \country{China}}
\email{wangying@swc.neu.edu.cn}

\author{Hai Yu}
\affiliation{%
  \institution{Northeastern University}
  \country{China}}
\email{yuhai@mail.neu.edu.cn}

\author{Na Meng}
\affiliation{%
  \institution{Virginia Tech}
  \country{USA}}
\email{nm8247@vt.edu}

\begin{abstract}
Ever since the launch of ChatGPT in 2022, a rising concern is whether \tool will replace programmers and kill jobs. Motivated by this widespread concern, we conducted an empirical study 
to systematically compare \tool against programmers in question-answering and software-maintaining. We reused a dataset introduced by prior work, which includes 130 StackOverflow (SO) discussion threads referred to by the Java developers of \repoCount GitHub projects. 
We mainly investigated three research questions (RQs). First, how does \tool compare with programmers when answering technical questions? Second, how do developers perceive the differences between \tool's answers and SO answers? Third, how does \tool compare with humans when revising code for maintenance requests? 

For RQ1, we provided the 130 SO questions to \tool, and manually compared \tool answers with the accepted/most popular SO answers in terms of relevance, readability, informativeness, comprehensiveness, and reusability. For RQ2,  
we conducted a user study with 30 developers, asking each developer to assess and compare 10 pairs of answers, without knowing the information source 
(i.e., \tool or SO).  
For RQ3, we distilled \taskCount software maintenance tasks from \taskCount GitHub projects citing the studied SO threads. We queried \tool to revise a given Java file, and to incorporate the code implementation for any prescribed maintenance requirement. 
Our study reveals interesting phenomena: For the majority of SO questions (\cBetter/130), \tool provided better answers; 
in 203 of 300 ratings, developers 
preferred \tool answers to SO answers; \tool revised code correctly for \taskGPT of the \taskCount tasks. Our research will expand people's knowledge of \tool capabilities, and shed light on future adoption of \tool by the software industry.

\end{abstract}

\ccsdesc[500]{Human-centered computing~Empirical studies in collaborative and social computing}
\ccsdesc[300]{Software and its engineering~Software maintenance tools}
\ccsdesc[100]{Software and its engineering~Search-based software engineering}

\keywords{empirical study, ChatGPT, StackOverflow, Q\&A, software maintenance}


\maketitle

\section{Introduction}

ChatGPT is a large language model-based chatbot developed by OpenAI; it can answer questions and assist users with different tasks, such as composing emails, essays, and code~\cite{chatgpt}. Ever since \tool's launch in November 2022, heated debates have spawned concerning its impacts on industries, society, economy, and regulations~\cite{chatgpt-the-wolf,chatgpt-destabilize,will-replace,gpt-took,within-10,will-replace-2,patel-post,will-replace-3,quora}. For instance, many people hold pessimistic attitudes~\cite{chatgpt-the-wolf,chatgpt-destabilize,will-replace,within-10}. They believe that \tool could replace white-collar workers in sectors like education, finance, software, journalism, and graphic design. 
 Meanwhile, some people are optimistic about \tool's role in the software industry, treating it as a coding assistant to help improve programmer productivity~\cite{will-replace-2,patel-post,will-replace-3,quora}. The reasons they mentioned include (1) ChatGPT uses existing code available online to answer questions, but has no creativity to produce new code~\cite{will-replace-2}; (2) it partially automates coding tasks~\cite{patel-post}; (3) it is only capable of basic coding~\cite{will-replace-3}. 
Albeit the heated discussion, little systematic study was done to qualitatively and quantitatively assess \tool's programming capability. 
 
To demystify the capability of \tool and predict its potential role in software development, we conducted an empirical study to thoroughly compare \tool against human programmers in two major scenarios: question answering and software maintenance. 
We chose these scenarios for two reasons. First, question-and-answer (Q\&A) is the main interaction mode between software developers and ChatGPT; the provided answers are essential in shaping the art and practices of software in the future. Thus, comparing the answers provided by both \tool and humans will help developers decide how to integrate \tool into their daily programming practices, and how much trust to give to \tool's answers. Second, software maintenance cost can make up to 90\% of the whole software development cost~\cite{software-maintenance}, which means that developers are likely to spend the majority of their time and effort maintaining software. By using \tool, developers can get answers instantly, without waiting for someone to notice their query on a forum. Therefore, our comparison between \tool's responses and developers' responses to the same maintenance requests will demonstrate how probably \tool can replace humans.

For our study, we reused a dataset introduced by prior work~\cite{stackoverflow-reuse}, which includes 130 StackOverflow (SO) discussion threads referenced by \repoCount GitHub projects. 
Each thread contains a question and at least an answer, with identifiable accepted or most popular answer(s). Each of the threads has URLs explicitly referenced by at least one GitHub project. Based on the dataset, we investigated the following three research questions (RQs):

\begin{itemize}
\item[\textbf{RQ1}] \emph{How does \tool compare with programmers when answering technical questions?} We provided 130 SO questions to \tool, and manually compared \tool answers with the accepted/most popular SO answers. The 130 samples belong to 5 major categories (e.g., optimization and debugging), and cover 9 technical topics (e.g., data processing and testing).
\item[\textbf{RQ2}] \emph{How do developers consider the differences between ChatGPT answers and SO answers?} We did a user study with 30 developers. We gave each participant 10 SO questions, 10 accepted or most popular SO answers, and 10 \tool answers. For each question, a participant assessed and compared the two given answers, without knowing the answer providers.
\item[\textbf{RQ3}] \emph{How does ChatGPT compare with humans when revising code for maintenance requests?} We mined the \repoCount GitHub repositories, for any commit that introduces both an SO link and code revision to an existing Java file. We formulated a prompt based on the SO thread and original Java file, asking \tool to produce a new version of that file to integrate the requested feature implementation. We formulated in total \taskCount prompts to query \tool. 
\end{itemize}

Our work provides empirical evidence for many interesting phenomena that are rarely mentioned by prior work. 
The major findings are summarized as below:

\begin{itemize}
\item For 75\% (\cBetter/130) of SO questions, \tool provided better answers than the accepted or most popular SO answers. 
The \cBetter answers respond to questions of all the styles and topics we examined. 
 It means that \tool is strong at answering various technical questions. 
\item Among the 300 ratings provided by developers, 68\% of ratings imply that \tool answers are better; only 32\% of ratings imply SO answers to be better. 
Compared with other question styles, 
\tool answers to comprehension questions were rated higher more often. 
\item Given \taskCount  maintenance tasks, \tool modified Java files for all tasks. However, only \taskGPT of these revised files can be smoothly integrated into GitHub repositories.
\tool-3.5 could not correctly maintain software in 54\% of cases. 
Developers need to do extra work to integrate or further improve the code revision recommended by \tool.
\end{itemize} 

\section{Background: The GitHub$\rightarrow$SO Dataset Used in Our Study}\label{sec:dataset}

In our research, one challenge is: \emph{What prompts should we provide to \tool, in order to compare it against developers in a realistic and fair manner?} To overcome this challenge, we chose to reuse the dataset constructed by Chen et al.~\cite{so-git} for their recent investigation on how GitHub developers reuse StackOverflow answers~\cite{stackoverflow-reuse}. The dataset consists of 130 StackOverflow (SO) discussion threads, referenced by Java files in \repoCount GitHub repositories. Every thread has a question post  and at least one answer post; every post is assigned with a unique URL.
Each of the included Java files references an SO thread via either the question URL, or URL of any answer belonging to that thread. 

\begin{figure}
\includegraphics[width=\linewidth]{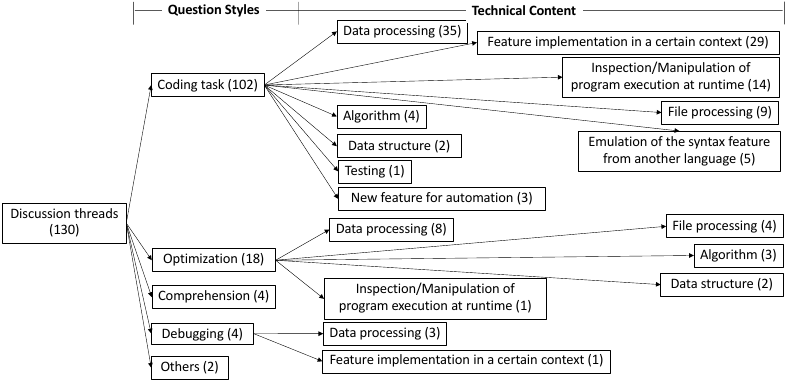}
\vspace{-2.em}
\caption{The taxonomy of SO threads based on both question styles and technical content~\cite{stackoverflow-reuse}}\label{fig:taxonomy}
\vspace{-1.2em}
\end{figure}
 
As shown in Fig.~\ref{fig:taxonomy}, Chen et al.~classified the 130 threads into 5 categories based on the question styles. Specifically, \emph{Coding task} means that askers describe software requirements and seek for code solutions. \emph{Optimization} means that askers provide
initial programs satisfying certain requirements, looking for
better programs that have either easier implementation, lower
runtime overheads, or less platform-specific dependency. 
\emph{Optimization} is different from the \emph{Coding task} category mentioned above,
as askers provide initial code implementation. \emph{Comprehension} is about clarification or comparison of concepts, terms, or APIs (e.g., \codefont{StringBuilder} vs.~\codefont{StringBuffer}). \emph{Debugging} means that askers present their erroneous programs, and solicit debugging feedback. \emph{Others} captures the miscellaneous
questions not covered by any category mentioned above. 
Additionally, Chen et al.~also classified the threads into nine topics based on the technical content.

According to Fig.~\ref{fig:taxonomy}, SO threads do not distribute evenly among different categories or topics. For example, \emph{Coding task} is the dominant category, covering 102 of the 130 threads. 
We intentionally chose this dataset instead of creating a balanced one, because this dataset reflects (i) the major concerns of developers when they discuss technical issues on SO, and (ii) developers' practices of software maintenance under the guidance of an online knowledge base.
Both (i) and (ii) can facilitate our evaluation of \tool, as they help us identify the most important questions to ask \tool, and provide good reference answers against which we can compare \tool's outputs. 


\section{Methodology}\label{sec:method}

There are three research questions (RQs) in our study: 

\begin{itemize}
\item[\textbf{RQ1}] \emph{How does \tool compare with programmers when answering technical questions?} This RQ examines given questions of different styles (e.g., coding or debugging task) or covering different topics (e.g., data structure or algorithm), whether \tool answers outperform or underperform SO answers.
\item[\textbf{RQ2}] \emph{How do developers consider the answer differences between \tool and SO?} This RQ assesses given \tool answers and SO answers, whether developers present obvious preferences towards one answer type.
\item[\textbf{RQ3}] \emph{How does ChatGPT compare with humans when revising code for maintenance?} 
This RQ explores \tool's capability in maintaining software, so it complements both RQ1 and RQ2 that examine \tool's capability in answering SO questions. 
\end{itemize}

\begin{figure}[h]
\begin{minipage}{.5\linewidth}
\includegraphics[width=\linewidth]{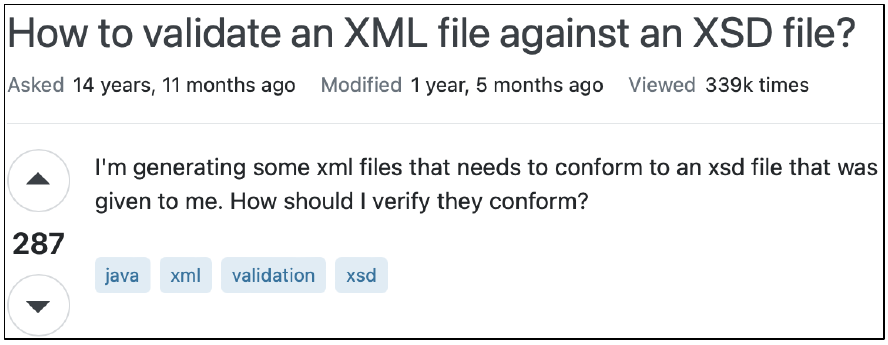}
\vspace{-2em}
\caption{An example SO question post}\label{fig:so-example}
\end{minipage}
\vspace{-1.em}
\hfill
\begin{minipage}{.49\linewidth}
\centering
\includegraphics[width=.95\linewidth]{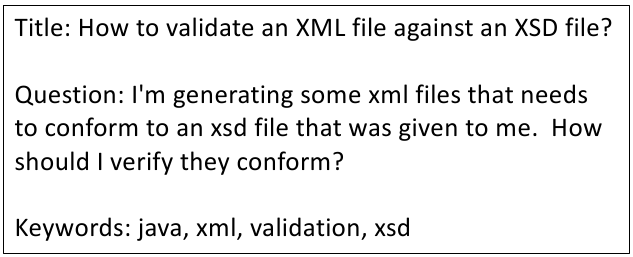}
\vspace{-.8em}
\caption{The prompt we crafted for the SO question}\label{fig:prompt-example}
\end{minipage}
\vspace{-1.em}
\end{figure}

\subsection{The Experiment Design for RQ1}
We located the most popular answer (i.e., the one with the highest vote) and accepted answer in each discussion thread, and considered such answers as the \emph{best answers provided by expert developers}. 
Notice that not every thread has an answer labeled as ``accepted'', while in many threads, the most popular answer is simultaneously the accepted answer. Therefore, we located 1-2 best answers in each thread. For simplicity, this paper refers to these answers as ``SO answers''. 
Additionally, we formulated a prompt for each SO question by extracting the title,  technical content, and keywords. For instance, given the SO question shown in Fig.~\ref{fig:so-example}, we crafted the prompt shown in Fig.~\ref{fig:prompt-example}, to ensure that \tool is exposed to the same amount of question information as humans. After sending 130 prompts to \tool, we collected all generated answers and manually compared those answers with SO answers. To avoid human bias by individual authors, the first two authors independently compared \tool answers with SO answers. For each group of answers under comparison, they rated which answer was better based on the following criteria:

\begin{enumerate}
\item Relevance. Does the answer directly respond to the question?
\item Readability. Does the answer clearly explain the solution?
\item Comprehensiveness. Is the solution comprehensive enough to cover all edge cases?
\item Informativeness. Does the answer contain code snippets to concretize the explanation?
\item Reusability. If a certain code is provided by the answer, is it easy to (re)use for developers?
\end{enumerate}
After rating all answer groups, we calculated Cohen's kappa coefficient value~\cite{kappa} to measure the inter-rater reliability. The measured value is 0.746, demonstrating a substantial agreement between the two authors. For the 14 ones on which they disagreed with each other, another author was invited to compare all answers and to lead a discussion until all three authors reached a consensus.

\subsection{The Experiment Design for RQ2}\label{sec:rq2-design}
We successfully got an IRB approval to conduct a user study, and recruited 30 Java developers by sending email invitations in our institution or spreading information via personal networks. Each participant accessed a Google form to compare \tool answers with SO answers for 10 SO questions, filled in his/her assessment, and submitted the form online. 
For all answers, we anonymized the sources: no participant knows which answer was generated by ChatGPT or SO.
Everyone spent 30--40 minutes to complete the survey, and got a 10-dollar gift card for compensation. 

One possible design of the user study could be asking all participants to compare the same 10 answer pairs. However, such a design can only check for people's opinions on 10 answer pairs at most, and considerably limit the generalizability of our research. Thus, we decided not to experiment in this way. Instead, 
from the dataset created for RQ1, 
we randomly selected \qCount SO questions, the corresponding best SO answers (i.e., either the accepted or most popular answer), and \tool answers. We ensured that the selected questions cover all five question styles mentioned in Section~\ref{sec:dataset}, and they are not too long to demotivate participants.  

As shown in Table~\ref{tab:study}, our selection includes 18 coding tasks, 7 optimization questions, 4 comprehension questions, 2 debugging questions, and 1 question belonging to the \emph{Other} category. 
We formulated 6 sets $[S_1, S_6]$ with these questions, so that each set has 10 SO questions to cover all 5 categories. We created six different google forms based on the six question sets, 
and sent five copies of each Google form to five participants.
To sum up, each question set is assessed by 5 participants; the 6 sets are assessed by 30 people and cover in total 32 questions, with some questions shared between sets because those questions are from smaller categories (e.g., \emph{Debugging} and \emph{Other}). 

For each pair of answers under comparison, developers responded to the following five questions:

\begin{table}
\scriptsize
\caption{The SO questions and participants included in our user study}\label{tab:study}\vspace{-1.5em}
\begin{tabular}{l r r }
\toprule
\textbf{Question Category} & \textbf{\# of questions} &\textbf{\# of participants per question} 
\\ \toprule
Coding Task & 18 & 5 \\
Optimization & 7 & 5--10 \\
Comprehension &4 &5--15\\
Debugging &2&30\\
Others &1&30
\\ \bottomrule
\textbf{Total} & 32 & 30 
\\ \bottomrule
\end{tabular}
\vspace{-1.5em}
\end{table}

\begin{itemize}
\item[Q1:] Is answer \#1 correct?
\item[Q2:] Is answer \#2 correct?
\item[Q3:] Does answer \#1 have better readability than answer \#2 (i.e., more readable)?
\item[Q4:] Does answer \#1 have better informativeness than answer \#2 (i.e., more informative)?
\item[Q5:] Considering all factors mentioned above, which answer do you prefer?
\end{itemize} 
Q1--Q2 are about the correctness checking for \tool answers and SO answers. Both questions have three options for developers to choose from: correct, unsure, and incorrect. Q3 and Q4 separately focus on the readability and informativeness of answers. Participants were expected to respond to both questions in a five-level Likert scale~\cite{allen2007likert}. Namely, Q3 and Q4 both have five options for developers to choose from: significantly better, slightly better, no difference, slightly worse, and significantly worse. Q5 is about developers' preference; it provides two options for participants to choose from: answer \#1 and answer \#2.

To avoid human bias due to sequential ordering, we randomized the order of questions and answers for each set.
 In scenarios where participants lacked the technical background to assess answers, we permitted them to search for relevant information online (e.g., articles or books). However, we asked them to intentionally avoid the information from SO, as the retrieved SO answers may reveal  identities of anonymized answers. We also asked them to avoid querying ChatGPT for answers, as those generated answers may  introduce bias towards \tool answers. 
Finally, we collected and analyzed the submitted Google forms.

\subsection{The Experiment Design for RQ3}
In the \repoCount GitHub repositories mentioned by Chen et al.~(see Section~\ref{sec:dataset}), we found 480 Java files to cite any of the 130 discussion threads. 
We believe that when a Java file was modified to include an SO discussion thread and code revision, 
it is very likely that developers modified the program in response to certain maintenance need and the cited thread captures that need. 
Based on this insight, we located $\langle$SO-thread, code-revision$\rangle$ pairs in Java files, and simulated maintenance tasks accordingly. 
By asking \tool to fulfill those tasks and by examining the tool's outputs, 
we assessed \tool's capability in code maintenance and its potential of replacing developers. 

\begin{figure}
\centering
\includegraphics[width=.92\linewidth]{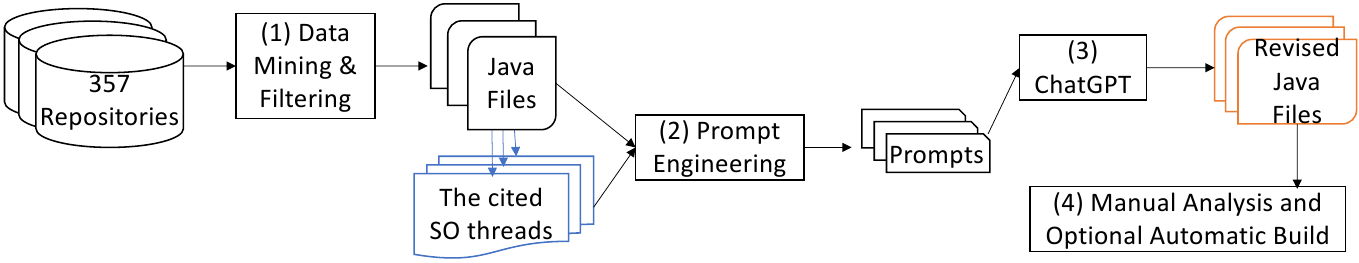}
\vspace{-1.em}
\caption{The experiment procedure for RQ3}\label{fig:procedure}
\vspace{-2.em}
\end{figure}

Fig.~\ref{fig:procedure} shows our four-step experiment procedure. Starting from the \repoCount repositories, we crawled the latest versions to identify 480 Java files that contain 501 SO references, with some files citing multiple SO references in one version or citing distinct SO references in different versions. 
We then refined the data by applying three filters.
First, we checked whether any of the Java files had the SO link added to its initial version. If so, we discarded the files because we could not easily locate the code region specifically relevant to the SO link.
Second, for any commit that adds an SO link to a Java file, we also checked whether any code was revised in that file by the same commit; if not, we discarded the file.
Third, for any file modified to include both an SO reference and code revision, we checked whether there is any semantic relevance between the two; 
if not, the file was discarded. 

In our study, the 3 filters separately removed  357, 29, and 10  SO references from the original 501  references. 
We had 105 Java files remaining, which were further divided
into 2 groups: 13 files from compilable projects and 92 files from uncompilable ones. 
Due to the time limit, we experimented with all 13  files from compilable projects and 35 sampled unique files from uncompilable projects. 
To simplify presentation, we use $f_i (i\in[1, 48])$ (i.e., 48 = 13 + 35) to refer to each file. 
All these files were modified in version history to introduce (1) references to SO discussions and (2) related code revisions. Thus, we use $f^o_i$ and $f^n_i$ to refer to the old and new versions of $f_i$. 


In Step 2, we formulated a prompt for $f_i$. The prompt
provides the content of $f^o_i$; it describes a maintenance request (e.g., feature addition or bug fixing) based on the SO question cited by $f^n_i$ as well as differences between $f^o_i$ and $f^n_i$. \tool is supposed 
to revise the given code $f^o_i$ in response to that request. 
Step 3 sends all prompts to ChatGPT. Step 4 gathers outputs from \tool and validates them via automatic build and/or manual inspection. For the first group of files (i.e., 13 files from compilable projects),  
this step 
replaces $f^n_i$ in each project with \tool's output $f^c_i$, checking whether the generated file is compilable or whether it is compatible with developers' code in other files. 
For the second group of files (i.e., 35 files from uncompilable projects), automatic build is inapplicable to validate \tool's outputs. For all \taskCount files revised by \tool, 
we manually compared $f^c_i$ with $f^n_i$ to examine the semantic equivalence. If $f^c_i$ matches $f^n_i$ in terms of the program logic or implementation algorithm, we consider \tool to succeed in the maintenance task.

\begin{figure}
\centering
\includegraphics[width=\linewidth]{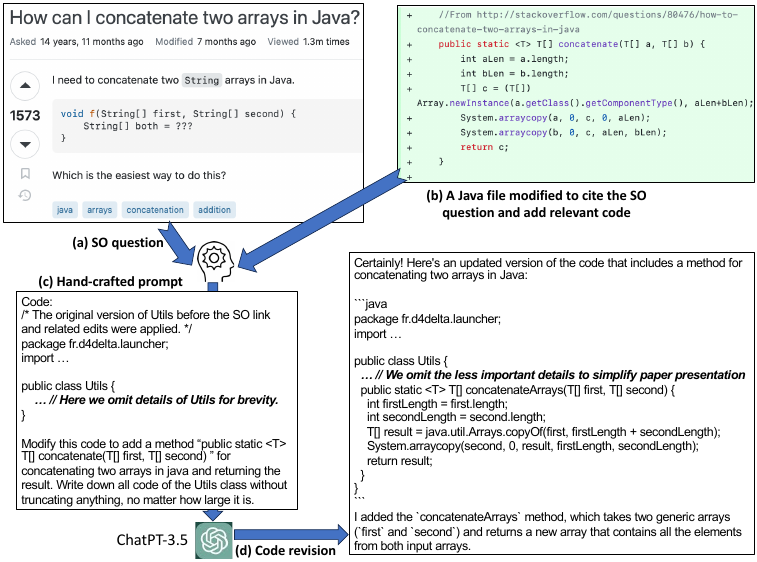}
\vspace{-2.em}
\caption{An example to illustrate our procedure of asking \tool to maintain a codebase}\label{fig:example}
\vspace{-1.em}
\end{figure}

Fig.~\ref{fig:example} shows a concrete example to illustrate the procedure.  
Our dataset has an SO question describing a coding task: to concatenate two arrays in Java (see Fig.~\ref{fig:example} (a)); there is also a GitHub repository with a Java file $f$ edited to cite the SO question and to add relevant code. For simplicity, Fig.~\ref{fig:example} (b) only presents the relevant diff data between program versions before and after edits. 
We inspected both the SO question and developers' related edits, to craft a prompt. As shown in Fig.~\ref{fig:example} (c), the prompt includes the original version of the Java file ($f^o$), and a request asking \tool to modify code and add a method for concatenating two arrays in Java. To ensure that \tool's output can be easily integrated into the GitHub project, we crafted the prompt to also (1) specify the method header of code-to-generate, and (2) require generating the full code of the revised file. 
Due to the space limit, Fig.~\ref{fig:example} (c) only illustrates a brief version of the actual Java code we included in the prompt. 
Afterwards, we sent \tool the prompt and got the modified file $f^c$ (see Fig.~\ref{fig:example} (d)). 
We then compared $f^c$ with developers' revision $f^n$, to assess \tool's maintenance capability.


\section{Experiment Results}\label{sec:result}
This section introduces and explains our experiment results. 

\subsection{Results for RQ1}\label{sec:rq1}
\begin{figure}
\includegraphics[width=\linewidth]{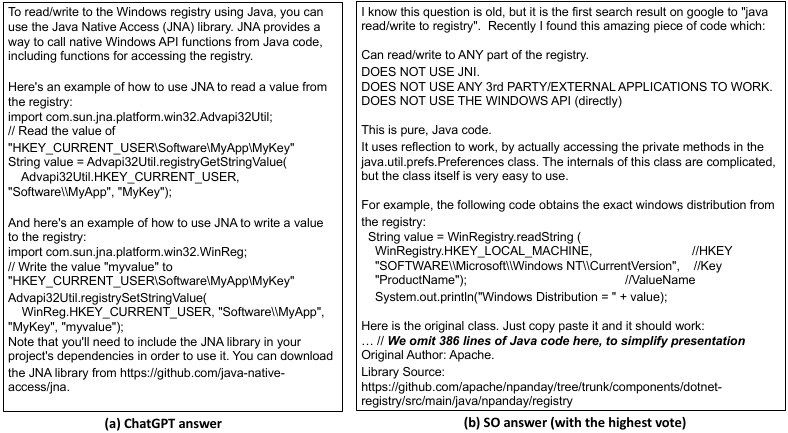}
\vspace{-2.2em}
\caption{Answers to the SO question ``How is it possible to read/write to the Windows registry using Java?''}\label{fig:answer-comparison}
\vspace{-1.em}
\end{figure}

Based on our manual analysis and comparison of all answers, we noticed that none of the SO/\tool answers is fundamentally wrong. Our answer preference was decided mainly based on the readability, comprehensiveness, and informativenes of descriptions. For example, Fig.~\ref{fig:answer-comparison} shows two answers to the SO question ``\emph{How is it possible to read/write to the Windows registry using Java?}'', including one \tool answer and the most popular SO answer. All authors agreed that the \tool answer is better, because it is more concise and clear: it explains how to read and write to the Windows registry using two simple code snippets. Meanwhile, the SO answer offers a 386-line code implementation copied from an open-source project, without explaining the essential internal program logic. Such answers may tempt developers to blindly copy-and-paste code or get into copyright issues, but does not necessarily help developers improve coding skills in the long run.

Among the 130 SO questions, the authors preferred \tool answers for \cBetter questions, and preferred SO answers for 33 questions. It means that \tool answers are often better. 
Table~\ref{tab:rq1} shows the result breakdown across five categories. According to this table, \tool provides better answers for 78\% of coding tasks, while SO answers are better for 22\% of coding tasks. For the four debugging questions, \tool answers are always better. Among optimization and other questions, \tool answers are better for 50\% of cases, while SO answers are better for the remaining 50\%. Among the four comprehension questions, \tool answers are better for three cases.

\begin{table}
\caption{Distribution of comparison results across question categories}\label{tab:rq1}
\vspace{-1.em}
\scriptsize
\begin{tabular}{l r r r r r}
\toprule
&\textbf{Coding Task} &\textbf{Optimization}&\textbf{Comprehension}&\textbf{Debugging} &\textbf{Other}\\ \toprule
\textbf{ChatGPT answer is better} &78\% (80/102)& 50\% (9/18) &75\% (3/4) &100\% (4/4) & 50\% (1/2)\\ 
\textbf{SO answer is better} & 22\% (22/102) & 50\% (9/18) & 25\% (1/4) &0\% (0/4) & 50\% (1/2)\\ 
\bottomrule
\end{tabular}
\vspace{-1.5em}
\end{table}

\begin{table}
\scriptsize
\caption{Distribution of comparison results across different 
subcategories of \emph{Coding Task} and \emph{Optimization}}\label{tab:rq1-2}
\vspace{-1.5em}
\begin{tabular}{l| p{5.4cm}| r| r}
\toprule
\textbf{Question Category} &\textbf{Technical Topic} & \textbf{ChatGPT answer is better} &\textbf{SO answer is better}\\ \toprule
\multirow{9}{*}{{Coding task}} &{Data processing} & 86\% (30/35) & 14\% (5/35)\\ \cline{2-4}
&Feature implementation in a certain context & 72\% (21/29) & 28\% (8/29) \\ \cline{2-4}
&Inspection/Manipulation of program execution at runtime & 64\% (9/14) & 36\% (5/14) \\ \cline{2-4}
& File processing & 89\% (8/9) &11\% (1/9)\\ \cline{2-4}
&Emulation of the syntax feature from another language & 60\% (3/5) & 40\% (2/5)\\ \cline{2-4}
& Algorithm & 75\% (3/4) & 25\% (1/4)\\ \cline{2-4}
& New feature for automation & 100\% (3/3) & 0\% (0/3) \\ \cline{2-4}
& Data structure & 100\% (2/2) & 0\% (0/2) \\ \cline{2-4}
& Testing & 100\% (1/1) & 0\% (0/1) \\ \hline

\multirow{5}{*}{Optimization} & Data processing & 62\% (5/8) &38\% (3/8) \\ \cline{2-4}
&File processing & 25\% (1/4)& 75\% (3/4) \\ \cline{2-4}
&Algorithm & 33\% (1/3) & 67\% (2/3) \\ \cline{2-4}
&Data structure & 50\% (1/2) & 50\% (1/2) \\ \bottomrule
\end{tabular}
\vspace{-1.em}
\end{table}

Due to the large number of data samples in categories \emph{Coding task} and \emph{Optimization}, we further zoomed into these categories to study how well the two types of answers compare with each other on different technical topics. As shown in Table~\ref{tab:rq1-2}, \tool answers generally outperform SO answers in all subcategories of \emph{Coding task}. For some minor topics like \emph{New feature for automation} (i.e., discussion on rare feature implementation), \emph{Data structure} (i.e., discussion on defining customized data structures), and \emph{Testing} (i.e., discussion on defining test cases), \tool answers are always better. 
One reason to explain our observations is that \tool answers
often have better clarity, readability, and/or larger coverage of edge cases and alternative solutions.

Interestingly, for optimization questions, \tool answers only outperform SO answers for 50\% of the cases. SO answers are better for the remaining 50\% of cases mainly because (1) the SO answers have more rigorous performance comparison between alternative code snippets, or (2) the SO answers suggest usage of advanced libraries instead of coding from scratch.

Compared with SO answers, \tool answers are often better because they (1) include as much relevant information as possible, and (2) provide more clear and concise explanations. 
However,  \tool answers may not outperform when 
askers look for optimized solutions. 


\noindent\begin{tabular}{|p{13.5cm}|}
	\hline
	\textbf{Finding 1:} \emph{For 75\% (\cBetter/130) of SO questions we studied, \tool answers are better than SO answers (i.e., accepted or most popular answers). For almost all the question styles and technical topics we examined, \tool answers are typically preferable or the dominant better answers.
	}
	\\
	\hline
\end{tabular}

\subsection{Results for RQ2}
\label{sec:rq2}

Table~\ref{tab:experience} presents the years of Java programming experience of the 30 participants in our user study. 
 As shown in the table, 20 developers have 1-2 years of experience, 8 developers have 3-5 years of experience, and 2 developers have 6-10 years of experience. $U_1$--$U_6$ denote the six user groups we created, based on the six question sets $S_1$--$S_6$ mentioned in Section~\ref{sec:rq2-design}. Namely, each group $U_i (i\in[1, 6])$ has five participants, assessing answers for $S_i$. As all survey questions anonymize the sources of answers under discussion, our analysis mapped answers to their sources (i.e., ChatGPT or SO) after the survey was done and before we derived the results discussed below.

\begin{table}[h]
\scriptsize
\caption{Years of Java programming experience of the 30 participants in our user study}\label{tab:experience}
\vspace{-1.5em}
\centering
\begin{tabular}{l| r r r r r r |r}
\toprule
 & \textbf{$U_1$} & \textbf{$U_2$} & \textbf{$U_3$} & \textbf{$U_4$} &\textbf{$U_5$} & \textbf{$U_6$} &\textbf{Total}\\ \toprule
 1-2 years & 1 & 4 & 4 & 3 & 4 & 4 & 20\\ 
 3-5 years &2 &	1&	1&	2&	1&	1&8 \\
 6-10 years & 2	 &0	&0	&0	&0	&0 &2\\
 >10 years & 0	 &0	&0	&0	&0	&0 &0\\ \bottomrule
\end{tabular}
\vspace{-1.em}
\end{table}


\subsubsection{Correctness Comparison}
As shown in Table~\ref{tab:correctness}, 
across all groups, ChatGPT answers received more ``Correct'' labels than SO answers (231 vs.~206), more ``Incorrect'' labels (29 vs.~25), but a lot fewer ``Unsure'' labels (40 vs.~69). 
For individual groups, ChatGPT answers received more ``Correct'' labels than SO answers in five groups, but received fewer ``Correct'' labels in $U_1$ only. ChatGPT answers received more ``Incorrect'' labels in four groups, but received fewer ``Incorrect'' labels in only two groups: $U_3$ and $U_5$. \tool answers received fewer ``Unsure'' labels in all groups. 

We also clustered developers' responses for each answer, and identified the assessments voted for by the majority. Namely, if N (N$\geq$5) developers  assessed the correctness of one answer, we identified the label commonly chosen by at least N/2 developers and treated it as the assessment voted for by the majority.
As shown in Table~\ref{tab:majority}, among the 32 sampled \tool answers, the majority voted for 30 correct answers, 1 unsure answer, and 1 incorrect answer. Both the incorrect and unsure answers are about coding tasks, probably because these tasks are hard to solve or the \tool answers are hard to evaluate. 
Among the 32 sampled SO answers, the majority voted for 25 correct answers and 3 unsure answers; 4 SO answers received no majority voting as developers' opinions diverge a lot. The three ``Unsure'' answers are all about coding tasks; the ``No majority'' answers separately correspond to one coding task, two debugging questions, and one comprehension question. 
Such phenomena imply that SO answers are generally harder to assess than \tool answers.


Our observations imply that compared with SO answers, \tool answers are more likely to be considered correct, and developers are more certain about their correctness assessment for \tool answers. 
This may be because \tool answers are more readable or more informative.


\begin{figure}
\scriptsize
\begin{minipage}{.53\linewidth}
\captionof{table}{Developers' assessments for answer correctness}\label{tab:correctness}
\vspace{-1.5em}
\begin{tabular}{p{1.5cm}| l| r r r r r r |r}
\toprule
\textbf{Answer Type} &\textbf{Label}&\textbf{$U_1$} & \textbf{$U_2$} & \textbf{$U_3$} & \textbf{$U_4$} &\textbf{$U_5$} & \textbf{$U_6$} &\textbf{Total}\\  \toprule
\multirow{3}{*}{\tool}& Correct&34&	40&	43&	31&	41&	42&	231\\
& Unsure &5&	3&	3&	14&	8&	7&	40\\ 
& Incorrect &11&	7&	4&	5&	1&	1&	29\\
\hline
\multirow{3}{*}{SO}& Correct&36&38	&32	&30&	36&	34&	206\\
& Unsure &9&	7&	10&	17&	10&	16	&69\\ 
& Incorrect &5&	5&	8&	3&	4&	0	&25\\ 
\bottomrule
\end{tabular}
\vspace{-1.em}
\end{minipage}
\hfill
\begin{minipage}{.4\linewidth}
\captionof{table}{The majority voting for answer correctness}\label{tab:majority}
\vspace{-1.5em}
\centering
\begin{tabular}{p{1.2cm}|r| r}
\toprule
& \textbf{\tool answers} &\textbf{SO answers}\\ \toprule
Correct & 30 & 25\\ \hline
Unsure & 1 &3\\  \hline
Incorrect & 1&0\\ \hline
No majority & 0 & 4\\
\bottomrule
\end{tabular}
\vspace{-1.em}
\end{minipage}
\end{figure}

\vspace{0.2em}
\noindent\begin{tabular}{|p{13.5cm}|}
	\hline
	\textbf{Finding 2:} \emph{Compared with SO answers, \tool answers were rated to be correct more often (231 vs.~206); developers were more certain when rating \tool answers.
}
	\\
	\hline
\end{tabular}

\begin{table}
\scriptsize
\caption{Developers' assessments for answer readability}\label{tab:readability}\vspace{-1.5em}
\begin{tabular}{l | r r r r r r | r}
\toprule
\textbf{Assessment} &$U_1$ &$U_2$ &$U_3$ &$U_4$ &$U_5$ &$U_6$ &\textbf{Total}\\ \toprule
ChatGPT answer is significantly more readable& 18&	13&	17&	16&	15&	8&	87\\
\tool answer is slightly more readable &10	&7&	10&	11&	16&	22&	76\\
\tool answer and SO answer are equally readable & 6&	11&	6&	13&9	&11&	56\\
SO answer is slightly more readable  & 11	&10&	10	&7	&7	&4&	49\\
SO answer is significantly more readable & 5	&9&	7&	3&	3	&5&	32\\ \bottomrule 
\end{tabular}\vspace{-1.em}
\end{table}

\begin{table}
\scriptsize
\caption{The clustering of readability assessments based on question categories}\label{tab:readability-3}
\vspace{-1.5em}
\begin{tabular}{l| R{1.2cm} r r r r}
\toprule
\textbf{Assessment} & \textbf{Coding Task} &\textbf{Optimization} &\textbf{Comprehension} & \textbf{Debugging} &\textbf{Others}\\ 
\toprule
ChatGPT answer is significantly more readable & 31\% &20\%&37\%& 20\% & 37\%\\
ChatGPT answer is slightly more readable & 24\% &30\%&30\% &27\% &20\%\\
\tool answer and SO answer are equally readable &17\% &28\%&12\% &13\% &27\%\\
SO answer is slightly more readable &14\%&13\%&17\% &22\%&13\%\\
SO answer is significantly more readable&13\%&8\%&5\% &18\%&3\%\\ \bottomrule
\end{tabular}
\vspace{-1.em}
\end{table}

\subsubsection{Readability Comparison}\label{sec:readability}
Table~\ref{tab:readability} presents developers' assessments of answer readability. Among the 300 ratings developers provided, 163 ratings (54\%) show that \tool answers are significantly or slightly more readable; 81 ratings (27\%) show that SO answers are more readable; and 56 ratings (19\%) indicate the equivalent readability between two types of answers. Furthermore, among the five rows in Table~\ref{tab:readability}, the first row ``\emph{\tool answer is significantly more readable}'' has the largest total count---87, while the last row ``\emph{SO answer is significantly more readable}'' has the smallest total count---32. In each of the six groups [$U_1$, $U_6$], there are always more \tool answers rated to have better readability than SO answers.

We also clustered developers' readability assessments based on question categories. As shown in Table~\ref{tab:readability-3}, the majority of ratings (>50\%) in three categories---\emph{Coding Task}, \emph{Comprehension}, and \emph{Others}---imply \tool answers to be more readable. 
Among all five categories, \emph{Comprehension} received the highest percentage of ratings for the higher readability of \tool answers---67\%, while \emph{Debugging} received the lowest---47\%.

\noindent\begin{tabular}{|p{13.5cm}|}
	\hline
	\textbf{Finding 3:} \emph{Developers tend to rate \tool answers to have better readability than SO answers. Compared with other answers, \tool answers to comprehension questions are most likely to be considered having better readability. 
}
	\\
	\hline
\end{tabular}

\subsubsection{Informativeness Comparison} 
Tables~\ref{tab:informativeness}--\ref{tab:informativeness-2} show developers' assessments of answer informativeness. 
These tables present similar phenomena to those reported for readability assessments (see Section~\ref{sec:readability}). For instance, as shown in Table~\ref{tab:informativeness}, among all assessments, 
the majority of ratings  (i.e., 180) shows that  
\tool answers are more informative, fewer ratings (i.e., 80) show 
SO answers to be more informative, and even fewer ratings (i.e., 40) indicate the equivalent informativeness between two types of answers. 
In Table~\ref{tab:informativeness-2}, the majority of ratings in all five categories imply that \tool answers are more informative. Compared with other categories, 
\tool answers to comprehension questions are most likely to have better informativeness.

\begin{table}
\scriptsize
\caption{The clustering of assessments for answer informativeness based on user groups}
\label{tab:informativeness}\vspace{-1.5em}
\begin{tabular}{l | r r r r r r | r}
\toprule
\textbf{Assessment} &$U_1$ &$U_2$ &$U_3$ &$U_4$ &$U_5$ &$U_6$ &\textbf{Total}\\ \toprule
ChatGPT answer is significantly more informative& 19&	14&21	&17&	18&	9	&98\\
\tool answer is slightly more informative &12	&9&	11&	12&	15&	23&	82\\
\tool answer and SO answer are equally informative  & 3 &	7&	5&	10&	7&	8&	40\\
SO answer is slightly more informative  & 10&	13&	8	&6	&8&	7	&52\\
SO answer is significantly more informative & 6&	7&	5&	5&	2&	3&	28\\ \bottomrule 

\end{tabular}
\vspace{-1.em}
\end{table}

\begin{table}
\scriptsize
\caption{The clustering of assessments for answer informativeness based on question categories}\label{tab:informativeness-2}\vspace{-1.5em}
\begin{tabular}{l| R{1.cm} r r r r}
\toprule
\textbf{Assessment} & \textbf{Coding Task} &\textbf{Optimization} &\textbf{Comprehension} & \textbf{Debugging} &\textbf{Others}\\ 
\toprule
ChatGPT answer is significantly more informative & 30\% &25\%&53\%& 27\% & 27\%\\
ChatGPT answer is slightly more informative & 28\% &28\%&27\% &27\% &30\%\\
\tool answer and SO answer are equally informative &13\% &17\%&10\% &13\% &13\%\\
SO answer is slightly more informative &18\%&20\%&8\% &23\%&17\%\\
SO answer is significantly more informative&11\%&10\%&2\% &10\%&13\%\\ \bottomrule
\end{tabular}
\vspace{-1.em}
\end{table}


We calculated the Pearson correlation coefficient~\cite{Benesty2009} between developers' assessments for readability and informativeness, to measure how strongly the two variables are related to each other. 
Specifically, we converted the five-level Likert scale to points, by having ``\emph{\tool answer is significantly more readable (or informative)}'' mapped to five points and getting ``\emph{SO answer is significantly more readable (or informative)}'' mapped to one point. 
This allows us to quantitatively represent developers' opinions with numbers, and to apply statistical analysis to the two groups of numeric values. 
{We found the Pearson correlation coefficient to be 0.73, with the p-value smaller than 0.00001. It means that the two variables are strongly correlated, and this correlation is statistically significant and real. 
 Namely, if a developer highly rates an answer's readability (or informativeness), he/she is likely to also highly rate the answer's informativeness (or readability). 
 }

\vspace{0.2em}
\noindent\begin{tabular}{|p{13.5cm}|}
	\hline
	\textbf{Finding 4:} \emph{Developers tend to consider \tool answers to be more informative. {There is a strong positive correlation between developers' assessments for readability and informativeness.}}
	\\
	\hline
\end{tabular}

\begin{table}
\scriptsize
\begin{minipage}{.5\linewidth}
      \caption{Preferences organized by groups}\label{tab:preference}\vspace{-1.5em}
      \centering
        \begin{tabular}{l| r| r}
\toprule
 \textbf{Group}& \textbf{\tool Chosen} &\textbf{SO Chosen}\\ \toprule
$U_1$& 66\% & 34\%  \\ 
$U_2$ &54\%  & 46\% \\
$U_3$ &66\%  & 34\% \\
$U_4$&72\% & 28\% \\
$U_5$&76\% &24\% \\
$U_6$ &72\% &28\%\\
 \bottomrule
\end{tabular}
\vspace{-1.em}
\end{minipage}%
\hfill
\begin{minipage}{.5\linewidth}
      \caption{Preferences organized by question categories}\label{tab:preference-2}\vspace{-1.5em}
      \centering
        \begin{tabular}{l| r| r}
\toprule
 \textbf{Question Category}& \textbf{\tool Chosen} &\textbf{SO Chosen}\\ \toprule
 Coding task &64\% & 36\% \\ 
 Optimization &65\% & 35\%\\
 Comprehension&83\% & 17\% \\
 Debugging&60\% & 40\%\\
 Other&67\% &33\%\\
 \bottomrule
  \textbf{Total} & 68\% & 32\%\\ \bottomrule
\end{tabular}
\vspace{-1.em}
\end{minipage}

\end{table}

\subsubsection{Developers' Overall Preferences}

Tables~\ref{tab:preference} and~\ref{tab:preference-2} show developers' overall answer preferences. In each user group, more developers preferred \tool answers to SO answers (see Table~\ref{tab:preference}). In 54\%--76\% of cases, developers preferred \tool answers. 
According to Table~\ref{tab:preference-2}, in each question category, \tool answers were chosen more often as the preferable ones.
In particular, for comprehension questions, the highest percentage of developers (83\%) prefer \tool answers over SO answers. It means that \tool is especially good at answering comprehension questions. 
In total, among the 300 ratings provided by developers, 203 ratings (68\%) were about their preferences for \tool answers, while only 97 ratings (32\%) show that developers preferred SO answers.  

Additionally, we clustered developers' responses for each answer pair under comparison, and identified the preferences voted for by the majority. Namely, if N (N$\geq$5) developers simultaneously compared an $\langle SO, \tool\rangle$ answer pair, we identified the label commonly chosen by at least N/2 developers and treated it as the preference voted for by the majority. As shown in Table~\ref{tab:preference-majority}, in 32 sampled answer pairs, the majority preferred 5 SO answers and 25 \tool answers.
The five SO answers respond to four coding tasks and one optimization question. 
In another two sampled answer pairs, responses are divided equally between SO and \tool as N is even. These answer pairs separately respond to a debugging question and an optimization question. Our results imply that SO answers sometimes outperform \tool answers when responding to coding tasks, debugging questions, and optimization requests.

\begin{figure}
\scriptsize
\begin{minipage}{.55\linewidth}
\centering
\captionof{table}{The Pearson correlation analysis between answer characteristics and developers' preferences}\label{tab:pearson}\vspace{-1.5em}
\begin{tabular}{l| r|r}
\toprule
\textbf{Variables} & \textbf{Coefficient} & \textbf{P-value} \\ \toprule
\tool Answer correctness vs.~Preferences & 0.26 &<0.00001 \\ \hline
SO Answer Correctness vs.~Preferences & -0.27 &<0.00001\\ \hline
Readability vs.~Preferences &0.7 &<0.00001\\ \hline
Informativeness vs.~Preferences &0.73& <0.00001\\ \hline
\end{tabular}
\vspace{-2em}
\end{minipage}
\hfill
\begin{minipage}{.38\linewidth}
\centering
\captionof{table}{The answer preference by the majority of developers}\label{tab:preference-majority}\vspace{-1.5em}
\begin{tabular}{p{2.cm}| R{2.3cm}}
\toprule
& \textbf{The majority voting}  \\ \toprule
(1) SO chosen & 5 \\ \hline
(2) \tool chosen & 25\\ \hline
Both (1) and (2) & 2 \\ \bottomrule
\end{tabular}
\vspace{-2.em}
\end{minipage}
\end{figure}

To better understand developers' answer preferences, 
we did statistical analysis to see whether developers' preferences are correlated with the correctness, readability, or informativeness of answers. 
Firstly, we mapped developers' 300 ratings for \tool answer correctness to numeric values: 3 (correct), 2 (unsure), and 1 (incorrect); we also mapped developers' 300 preference responses to numeric values: 1 (SO answer), and 2 (\tool answer). We then applied the Pearson correlation analysis to the two groups of numeric data. 
As shown in Table~\ref{tab:pearson}, the coefficient is 0.26, implying that developers' preferences are weakly related to the correctness of \tool answers. Secondly, we repeated the above-mentioned process to study the correlation between developers' preferences and the correctness of SO answers. As shown in Table~\ref{tab:pearson}, the two variables are also weakly related. 

Furthermore, we adopted the two-way ANOVA test~\cite{Yates1934}---a statistical method to examine the influence of two different categorical independent variables on one continuous dependent variable. By applying this method, we examined whether developers' preferences depend on the correctness contrast  
within the $\langle SO, \tool\rangle$ answer pairs for given questions. Our analysis shows all evaluated p-values to be greater than 0.05, which means that the correctness contrasts between SO answers and \tool answers do not determine developers' preferences. All these three statistical tests imply that 
developers did not base their preference decisions on  
answer correctness. One possible reason is that the answers-under-comparison are often correct, or have little difference in terms of the correctness property.

Next, we applied Pearson correlation analysis to (1) developers' ratings of readability and preferences, and (2) developers' ratings of informativeness and preferences. As shown in Table~\ref{tab:pearson}, both tests produced high coefficient values: 0.7 and 0.73, and low p-values (<0.00001). The phenomena indicate that developers expressed their preferences mainly based on the readability and informativeness of answers.

\vspace{0.2em}
\noindent\begin{tabular}{|p{13.5cm}|}
	\hline
	\textbf{Finding 5:} \emph{Among the 300 ratings provided by developers, 203 ratings show that developers preferred \tool answers. Our statistical analysis shows that developers' preferences were mainly affected by the answer readability and informativeness, but not by the answer correctness.
	}
	\\
	\hline
\end{tabular}
\vspace{0.2em}

Finally, we compared developers' manual analysis results against ours described in Section~\ref{sec:rq1}.  Our answer preferences match developers' preferences in 84\% (27/32) of cases. There are only five cases where our preferences do not match developers'. One of the cases is about optimization, and we are very confident that SO provides a more efficient code solution than \tool. The other four cases cover three coding tasks and one comprehension question; the preference divergence is mainly due to personal styles or coding habits.
\subsection{Results for RQ3}\label{sec:rq3}
This section reports our experiments with the tasks separately defined for \taskAuto compilable projects and \taskManual uncompilable ones.

\subsubsection{Experiment with the \taskAuto tasks defined for  compilable projects.}
As shown in Table~\ref{tab:maintenance},
our prompts ask \tool to (1) add one or more methods to an existing Java file, (2) modify a return statement in an existing method, or (3) revise the implementation of existing Java methods.
 ChatGPT was able to generate revised Java files for all prompts. 
By trying to compile the files output by ChatGPT, we found 11 out of the 13 files to compile successfully. One file (see M1) does not compile, because \tool omitted almost all details of unchanged code in the given Java file and it majorly presented the added code implementation. Another file (see M13) does not compile because \tool introduced the usage of a variable \codefont{radius}, without defining or declaring \codefont{radius} first.

\begin{table}
\scriptsize
\caption{The \taskAuto maintenance tasks we created in compilable projects for \tool to fulfill}\label{tab:maintenance}
\vspace{-1.5em}
\begin{tabular}{r| p{2.2cm}| p{4.cm}| p{1.8cm}|p{3.6cm}}
\toprule
\textbf{Id} &\textbf{Program} &\textbf{Maintenance task} &\textbf{Does \tool's code compile?} &\textbf{Does \tool's code semantically match developers' code?} \\ \toprule
M1 & LOFiles~\cite{LOFiles} & Add a Java method to clone an entire document using the Java DOM.&No& No. \tool omits details of unchanged code. 
\\ \hline
M2 &CoreNLP~\cite{CoreNLP} & Add a method to concatenate two arrays in Java and return the result. & Yes & Yes. \tool's code also adds extra sanity checks for inputs.\\ \hline
M3 & DeltaLauncher~\cite{DeltaLauncher} & Add a method to concatenate two arrays in Java and return the result. & Yes & Yes  \\\hline
M4 &jlib~\cite{jlib} & Add a method to check whether the char-typed input parameter is printable. &Yes & No. \tool's code considers fewer corner cases.\\\hline
M5 &Achilles~\cite{Achilles} & Add a method to programmatically determine the availability of a port in a given machine. & Yes & No. Divergent values are assigned to the same field; \tool's code does not use the parameter {\tt hostname}.\\\hline
M6 & lanterna~\cite{lanterna} &Modify a return-statement, to make sure the returned value is true when the given character is printable.&Yes & No. \tool's code considers fewer corner cases.\\ \hline
M7 & gnikrap~\cite{gnikrap} & Add two methods to separately decode and encode base64 data.&Yes &Yes \\ \hline
M8 &the-holy-braille~\cite{the-holy-braille} &Add a method to count the lines of a file &Yes & Yes\\\hline
M9 &Aiolos~\cite{Aiolos} &Modify the code to configure a logger of the type java.util.logging.Logger, to have Level.ALL.&Yes  & No. Developers' code contains more project-specific logic.\\ \hline
M10 &CodingProblems~\cite{CodingProblems} &Add a method to implement an optimized algorithm of checking if an integer's square root is an integer. &Yes & No. \tool's code is not an optimized solution.\\ \hline
M11 &markov-test~\cite{markov-test} & Revise an existing method, so that it reads the content of a specified file, creates a Java string from that content, and returns the value. &Yes & Yes\\ \hline
M12 &opennars~\cite{opennars} &Add a method to return the exponential value of a given input, in a optimized and fast way.&Yes & No. ChatGPT omits details of unchanged code.\\\hline
M13 &skyroad-magnets~\cite{skyroad-magnets} & Modify an existing method, to check whether the circle and rectangle intersect in 2D euclidean space.&No & No. The variable {\tt radius} is used but not defined; the output value is calculated differently.\\\hline
\end{tabular}
\vspace{-1.5em}
\end{table}

For only 5 out of the 13 cases, \tool successfully revised given files to satisfy maintenance needs.
The prompts of all these five cases perfectly match the program logic implemented in developers' code, and the logic is totally irrelevant to the surrounding program context or other methods/classes defined in the same project. 
Specifically for M2, \tool not only output a correct Java file, but also added extra checks for the input parameters to avoid null-pointer dereferences.

\begin{figure}[h]
\centering
\includegraphics[width=\linewidth]{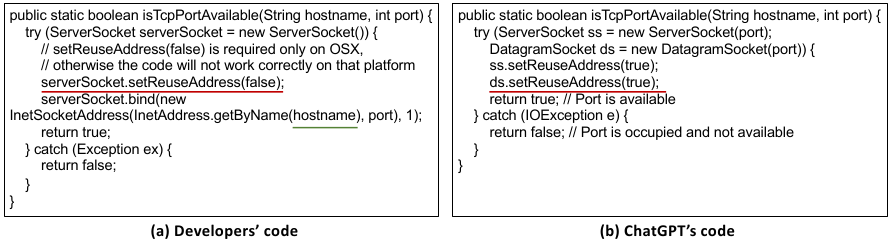}
\vspace{-2.em}
\caption{\tool's code does not match developers' code for M5, as it allows for reuse of the port
}\label{fig:developer-chatgpt}\vspace{-1.5em}
\end{figure}

For 8 out of the 13 cases, \tool did not revise given files as expected. In addition to the compilation issue mentioned above, two major reasons explain our observations. First, \tool's code handles fewer corner cases than developers' code (see M4 and M6). For instance, M6 requires for a method addition to check whether a given \codefont{char}-typed variable is printable. \tool's code considers limited types of non-printable characters (i.e., line separator, paragraph separator, and unassigned characters). However, developers' code covers more types of non-printable characters (e.g., keyboard characters like ``Tab''). 
Second, \tool could not generate project-specific logic (see M5, M9, M13). For instance, Fig.~\ref{fig:developer-chatgpt} presents the implementations by both developers and \tool for M5. Both snippets satisfy the requirement of determining the availability of a port in a machine. However, developers' code calls \codefont{setReuseAddress(...)} with the \codefont{false} parameter value, while \tool's code calls that method with \codefont{true}; developers' code uses the input \codefont{hostname} but \tool's does not. 
Although \tool's code is reasonable; it does not fit the program context.

\vspace{0.2em}
\noindent\begin{tabular}{|p{13.5cm}|}
	\hline
	\textbf{Finding 6:} \emph{Among the \taskAuto maintenance tasks defined for compilable projects, \tool only successfully fulfilled 5 tasks. This implies that \tool usually cannot maintain software independently.}
	\\
	\hline
\end{tabular}

\begin{table}[h]
\scriptsize
\caption{The \taskManual maintenance tasks we created in uncompilable projects for \tool to fulfill}\label{tab:developer-chatgpt-2}\vspace{-1.5em}
\begin{tabular}{l| p{5.8cm} ? l| p{5.8cm}}
\toprule
\textbf{Id} &\textbf{Does \tool's code semantically match developers' code?} & \textbf{Id} &\textbf{Does \tool's code semantically match developers' code?} \\ \toprule

M14 & Yes &M32 & No. \tool's code is incomplete.\\ \hline
M15 & No. \tool's code is not an optimized solution, as required. & M33 &Yes.\\ \hline
M16 & Yes & M34 &Yes. \\ \hline
M17 & No. \tool's code does not contain project-specific logic. & M35 &No. \tool's code is incomplete.\\ \hline
M18 & No. \tool's code is incomplete, and it also misses project-specific logic. &M36 &No. \tool's code calls an API that can throw an exception, but the method header does not declare that exception-to-throw. Meanwhile, developers' code does not call any API that can throw exception(s).\\ \hline
M19 & Yes & M37 & No. \tool's code can throw an exception, but the developers' code does not throw any exception. 
\\ \hline
M20 &No. \tool's code can throw an exception, but the developers' code does not throw any exception. 
&M38& Yes \\ \hline
M21 & No. \tool's code misses project-specific logic. 
&M39 & 
No. \tool's code can throw an exception, developers' code does not throw any exception. \\ \hline
M22 & Yes & M40 &Yes\\ \hline
M23 & Yes &M41 &Yes\\ \hline

M24 & No. \tool's code misses project-specific logic.&M42 &No. \tool's code covers fewer corner cases. \\ \hline
M25 &No. \tool's code covers fewer corner cases. &M43 &No. \tool's code fails to use one of the input parameters.\\ \hline
M26 & Yes. \tool's code further does sanity checks for inputs.&M44&Yes \\ \hline
M27 & No. \tool's code does not match project-specific logic. &M45&No. \tool's code is incomplete.\\ \hline
M28 & No. \tool's code covers fewer corner cases. &M46& No. \tool's code is incomplete.\\ \hline
M29 & Yes & M47 &Yes\\ \hline
M30 & Yes &M48 &Yes\\ \hline
M31 & Yes \\ \hline
\end{tabular}
\vspace{-1.em}
\end{table}

\begin{figure}
\centering
\includegraphics[width=.83\linewidth]{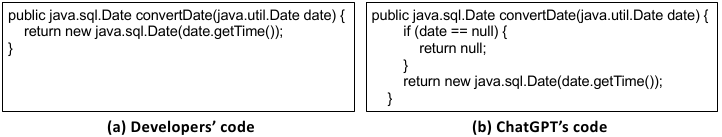}
\vspace{-1em}
\caption{\tool's code matches developers' code for M26, and conducts an extra sanity check for the input}\label{fig:developer-chatgpt-2}\vspace{-1.5em}
\end{figure}

\subsubsection{Experiment with the \taskManual tasks defined for uncompilable projects}
As shown in Table~\ref{tab:developer-chatgpt-2}, given the \taskManual tasks, \tool always output code to respond to our prompts. Among the responses, 17 match the program logic in developers' code and 18 responses do not. For 1 of the 17 matching cases---M26, \tool's code performs an extra check on the input parameter to eliminate potential program errors. As shown in Fig.~\ref{fig:developer-chatgpt-2}, both implementations satisfy the maintenance need of ``adding a method to convert \codefont{java.util.Date} to \codefont{java.sql.Date}''. However, \tool's version is safer, as it performs a null-pointer check before dereferencing the object \codefont{date}.

We identified 4 major reasons to explain why \tool's code failed to match developers' code in 18 cases. 
First, for 10 cases, \tool's code lacks the project-specific logic even though it satisfies the described maintenance need. Namely, \tool's code either (1) contains \codefont{throw}-statements to throw exceptions that are not thrown by developers' code, (2) misses some \codefont{if}-condition checks or other statements, or (3) fails to use all input parameters. 
Second, for five cases, \tool's code omits details of the unchanged code, and only outputs the revised code with limited surrounding context (i.e., unchanged code). Third, for three cases, \tool's code considers fewer corner cases. Fourth, for one case, \tool's code does not implement any optimized algorithm to decide whether an integer is a perfect square, although the prompts ask for such an optimization. 

\vspace{0.2em}
\noindent\begin{tabular}{|p{13.5cm}|}
	\hline
	\textbf{Finding 7:} \emph{Among the \taskManual maintenance tasks defined for uncompilable projects, \tool could not fulfill 18 tasks mainly because the generated code is either incomplete, lacking project-specific logic, covering fewer corner cases, or failing to use all input parameters.}
	\\
	\hline
\end{tabular}

\section{Threats to Validity}

\paragraph*{Threats to External Validity: } All our observations are limited to the SO discussion threads, GitHub repositories, and developers' responses included into our dataset. They may not generalize well to other SO discussion threads, other repositories (including closed-source repositories), or other developers. 
Our study focuses on Java programs, as the language has been popularly used and we have more experience and domain knowledge relevant to it; we did not apply \tool to programs written in other languages. 
We experimented with \tool-3.5 instead of \tool-4.0, because \tool-4.0 is not free to use. People may observe slightly different phenomena when applying higher versions of \tool to programs written in other languages.
In the future, to make our research findings more representative, we will expand our dataset and explore to use \tool-4.0.

\vspace{.3em}
\emph{Threats to Internal Validity: } We experimented with the default setting of ChatGPT-3.5, without controlling or tuning any parameter it defines. By default, when ChatGPT is queried with the same prompt multiple times, it generates results with randomness, i.e., it can produce different results given the same prompt. Such randomness can impact the validity or certainty of our observations.
However, based on our experience so far, ChatGPT often produces very similar results given multiple trials of the same prompt. We believe that the internal randomness of ChatGPT does not significantly impact our experiment results.

\vspace{.3em}
\emph{Threats to Construct Validity: } The manual inspection of SO answers and Java code is subject to human bias. To mitigate the potential inaccuracy due to human bias, for RQ1, we had two authors separately compare 130 pairs of $\langle SO, \tool\rangle$ answers; to resolve any opinion conflict between them, another author lead a discussion until the  three authors  reached a consensus. For RQ2, we recruited as many developers as possible (i.e., 30 developers), asked each to compare 10 answer pairs, and ensured each answer pair to be inspected by at least 5 developers. For RQ3, we leveraged automatic compilation to validate \tool's outputs whenever possible, and had three authors inspect \tool's outputs and the optional compilation reports. 

ChatGPT was trained on large collections of text data (e.g., books,
articles, and web pages) publicly available by September 2021. All data used in our study was available before that date. Therefore, model-overfitting issues may occur and our evaluation may overestimate \tool's capabilities. However, considering the large volume of training data used by \tool (i.e., 570 GB) and potential interference/conflicts among that data, we feel the overfitting issues to be insignificant.
 In the future, we plan to evaluate \tool with more recent data to mitigate this threat.
\section{Lessons Learned}

Below are the lessons or actionable items we learned from this study.

\paragraph{\textbf{For Developers or Potential \tool Users:}} 
\emph{\tool is a good information resource to refer to when developers have technical questions.} 
We observed that 
{{\tool answered SO questions with high accuracy, regardless of the question style or technical topic.}} According to ChatGPT itself, ``I was pre-trained using a combination of unsupervised and supervised learning techniques, such as language modeling, auto-encoding, and sequence prediction.''~\cite{chatgpt-work}. The pre-trained knowledge enables \tool to produce answers to SO questions with high accuracy. 
Another important aspect that has worked in ChatGPT's advantage is how quickly it responds. Developers can get answers instantly, without waiting for someone to notice their query on a forum. This real-time interaction allows developers to maintain their momentum and tackle coding challenges without unnecessary delays.
Among the five question categories we studied, \tool answers are generally better than SO answers for coding tasks and comprehension questions, but competitive for optimization tasks. 

\emph{Developers need to be very
cautious when merging ChatGPT’s code into their own projects.} Compared with developers’ code, ChatGPT’s code often does not fit into the context of given
Java files, even if it can satisfy the maintenance needs. Such a limitation is due to the unchanged code omitted by \tool, the missing program-specific logic, or fewer corner cases covered by \tool's code. The lack of project-specific logic may be caused by ambiguous
or narrow descriptions of prompts, or by ChatGPT’s limited capability of code generation.
To reasonably simulate real-developers' efforts, we did not further refine prompts to describe all project-specific requirements
in addition to the basic maintenance need. It is possible that \tool can work better when more project-specific details are specified in prompts; however, further investigation is still needed to validate this argument.

\vspace{0.3em}
\emph{\textbf{For SE Community and Q\&A Forums:}} \emph{When people create documentation (e.g., coding answers or user manuals) to provide guidance on software development or usage, they can consider using \tool to improve the readability and informativeness of documents.} Compared with the accepted or most popular SO answers provided by human experts (e.g., experienced developers), \tool's answers are often more desirable. Developers typically prefer \tool's answers due to the better readability and higher informativeness, but not necessarily due to the correctness. 

Our observations also raise a potential challenge for developer websites such as Stack Overflow and the whole SE community. Users may use these Q\&A websites less frequently because they can get better answers from ChatGPT, and even users of these sites may start using ChatGPT-like tools to formulate and post their answers.
In the long run, these behaviors may reduce the human-crafted materials available on the Internet for training.
\textit{It is unknown and worth further investigation whether such a reduction trend will negatively impact \tool-like tools.}


\vspace{.3em}
\textbf{\emph{For SE Researchers:}}
\emph{New program analysis techniques can be invented to identify the scenarios where it is safer to merge \tool's code into projects. Novel testing techniques can be created to specifically test the interaction between \tool's code and developers'  code in the same project.} In our study, \tool performed well on certain types of maintenance tasks, especially when it generated elementary functions to implement independent features and those features have little or no data dependency on other parts of the same project. Characterizing the scenarios when \tool's code is easy to (re)use can help researchers better assess the automation boundary of \tool, and help developers better leverage \tool to improve programmer productivity and software quality.

So far, we have not observed any obvious or dummy error in the outputs by \tool to signal significant limitations of the tool, or to facilitate our immediate rejection of specific answers.
Consequently, we always spent lots of time reading every single line of the outputs, thought carefully, and discussed thoroughly to identify issues lying in those outputs. 
If developers are not careful enough or are under time pressure, they may get misled by \tool and blindly accept all of the tool's outputs. \textit{To avoid the potential hallucination problem or misinformation produced by \tool, researchers need to apply comprehensive program analysis tools or even invent new tools to examine the tool's outputs rigorously.}

\vspace{.3em}
\emph{\textbf{For LLM Researchers:} Research needs to be done to investigate the best user interfaces that facilitate users to provide all necessary information for accurate code generation or appropriate program revision.} 
 One big challenge developers or LLM users would face is that it is unknown (1) how much project-specific information is sufficient, and (2) what project-specific information will help. Due to the size of code base and intellectual property restrictions, it is unlikely that developers can use their whole code base as the prompt. 
Meanwhile, for developers, the difficulty of describing all project-specific details can be sometimes equivalent to or even higher than that of writing code manually. \emph{It is also worth investigation how willing 
developers are to specify all project-specific details for coding tasks. The cost-efficiency of using \tool forms another important research direction}. 


\section{Related Work}
The related work of our research includes empirical studies on SO posts, and studies on \tool.  

\subsection{Empirical Studies on StackOverflow}
Researchers did various studies to characterize the crowdsourced knowledge available on StackOverflow~\cite{Nasehi2012,Movshovitz-Attias2013,Pinto2014,Gantayat2015,Honsel2015,Zou2015,Fischer2017,Baltes2017,meng2018secure,Zhang2018,Meng2019:icse,Bangash2019,Openja2020,Zahedi20,Firouzi2020,Zhang2021,Peruma:2021aa,Zhu:2022aa,Lu2022,Shoaibi2023}. 
Specifically, 
Zhang et al.~\cite{Zhang2021} explored how the knowledge in answers becomes obsolete. 
Gantayat et al.~\cite{Gantayat2015} studied the synergy between voting and acceptance of answers on SO. 
Zhu et al.~\cite{Zhu:2022aa} analyzed the question discussions on SO.
Some researchers studied the behaviors and activities of developers relevant to SO. For instance, Movshovitz-Attias et al.~\cite{Movshovitz-Attias2013} analyzed the SO reputation system to identify the participation patterns of high- and low-reputation users.  
Baltes et al.~\cite{Baltes2017} analyzed the attributed and unattributed usages of SO code snippets in GitHub projects. 
Lu et al.~\cite{Lu2022} conducted a survey with 938 developers who participate in SO to understand their participation motivations and incentive perceptions.

Some researchers examined SO threads related to specialized domains or technical topics.
For instance, 
Pinto et al.~\cite{Pinto2014} analyzed energy consumption-related posts to explore developers' concerns, the critical aspects of energy consumption, and developers' solutions to improve energy efficiency. 
Shoaibi et al.~\cite{Shoaibi2023} conducted an exploratory study on the challenges developers face in resolving performance regression. 
Peruma et al.~\cite{Peruma:2021aa} analyzed refactoring discussions on SO, and revealed five areas where developers typically require help: Code Optimization, Tools and IDEs, Architecture and Design Patterns, Unit Testing, and Database. 
Zahedi et al.~\cite{Zahedi20} applied topic modeling to SO posts to understand the non-functional requirements (NFRs) that developers focus on.
Firouzi et al.~\cite{Firouzi2020} studied 2,283 C\# snippets mined in SO data dump, to investigate why developers used unsafe codes (i.e., code blocks encapsulated via the C\# unsafe keyword).

Zhang et al.~\cite{Zhang2018} studied the code examples of API usage, to reveal answers that may misuse APIs.
Bangash et al.~\cite{Bangash2019} analyzed the machine learning-related posts, to investigate developers' understanding of machine learning. 
Openja et al.~\cite{Openja2020} analyzed release engineering questions, to understand the modern release engineering topics of interest and their difficulty.  
Fischer et al.~\cite{Fischer2017} and Meng et al.~\cite{meng2018secure,Meng2019:icse} examined SO posts related to Java security, to identify developers' concerns on security implementation, technical challenges, or vulnerabilities in answer code.

Our study complements all studies mentioned above, but has a unique focus on the comparison between SO answers and \tool answers.
No existing work compares \tool answers with SO answers.
 We mainly focused on the best SO answers developers could provide, including accepted and most popular answers. By comparing these answers with \tool answers, we intended to reveal how \tool compares with human experts in responding to technical questions. 


\subsection{Studies on ChatGPT}

Several studies were conducted on ChatGPT~\cite{Nascimento2023,tian2023chatgpt,Jalil2023,sobania2023analysis,Nikolaidis2023,chen2023gptutor}. Specifically, Nascimento et al.~\cite{Nascimento2023} used four LeetCode questions to create prompts, and observed \tool to outperform novice developers in solving easy or medium problems. Jalil et al.~\cite{Jalil2023} checked how well \tool performs when tasked with answering common questions in a popular curriculum, and found \tool to respond to 77.5\% of questions. 
Sobania et al.~\cite{sobania2023analysis} used a bug fixing benchmark set---QuixBugs; they found \tool to fix 31 out of 40 bugs and outperform the state-of-the-art. 
Tian et al.~\cite{tian2023chatgpt} assessed \tool's capability in code generation, program repair, and code summarization. 
They observed \tool to outperform two large language models 
in code generation; it is competitive with a state-of-the-art repair tool; it produces consistent summaries for code with the same intention.  
Nikolaidis et al.~\cite{Nikolaidis2023} evaluated \tool and Copilot using LeetCode problems. They found both models to well solve easy problems. 
Chen et al.~\cite{chen2023gptutor} created GPTutor, a \tool-powered programming tool, to provide code explanation for developers in IDE. 

Our study complements all prior work, as we studied \tool from unique angles. We characterized its capability of (1) answering SO questions and (2) maintaining or revising software in response to new software requirements. We further examined developers' opinions on the comparison between \tool answers and best SO answers. 

\section{Conclusion}

Motivated by the widespread concern on \tool's capability of replacing developers and killing jobs, 
we explored to use \tool in two typical working scenarios in developers' daily lives: question answering and software maintenance. 
We hypothesized that \tool could not provide good answers to technical questions, or satisfy the maintenance needs in given software projects. Surprisingly, we observed \tool to work very well in answering technical questions, and provide promising outputs to facilitate software maintenance. Specifically, 
both our manual inspection and user study show that 
given technical questions, \tool answers are often correct and reasonable; they often have higher quality than the most popular or accepted answers from SO. This implies that developers can always refer to \tool as a reliable information resource when they have technical questions; answer-providers or technical supporters can also leverage \tool to polish or enhance their original answers, to better help other developers, and to better shape the art as well as practice of software today and in future.

Meanwhile, \tool's responses to maintenance tasks are less satisfactory; the code included in these responses do not fit into the given program contexts in most cases either due to (1) \tool's limited understanding of program context, (2) its limited capability of code generation, or (3) the unclearness or ambiguity in task-describing prompts. We do not consider \tool to be able to replace humans or work as independent software maintainers, although we do observe
\tool's great capability in generating independent functional units (e.g., Java classes or methods) that have no or little dependency on the surrounding program context. 

To sum up, we are cautiously optimistic about \tool's role in the software industry. By quantitatively and qualitatively measuring its capabilities in question-answering and software-maintaining, our study characterizes the potential technical support and automation opportunities \tool brings; our study also reveals the potential pitfalls or challenges provoked by the tool. In the future, we will create better tools to automatically assess the quality of \tool's outputs, or integrate \tool into the existing tool chains for test generation or bug detection.


\bibliographystyle{ACM-Reference-Format}
\bibliography{mahir-conference-2024}

\end{document}